\documentclass[twocolumn,
aps,nofootinbib,showpacs,showkeys,preprint
tightenlines
] {revtex4-1}

\usepackage{epsf,epsfig,subfigure,graphicx,amsmath,amssymb}
\usepackage{color}

\newcommand{\dis}[1]{\begin{equation}\begin{split}#1\end{split}\end{equation}}

\newcommand{\gev}{\,\textrm{GeV}}

\newcommand{\ie}{{\it i.e.}\ }

\begin{document}

\title{\Large\bf The D0 same-charge dimuon asymmetry and possibile new CP violation sources in the $B_s-\overline{B}_s$ system}

\author{Jihn E. Kim\email{jekim@ctp.snu.ac.kr},  Min-Seok Seo\email{minseokseo@phya.snu.ac.kr} and  Seodong Shin\email{sshin@phya.snu.ac.kr}}
\affiliation{ Department of Physics and Astronomy and Center for Theoretical Physics, Seoul National University, Seoul 151-747, Korea
 }
\begin{abstract}
Recently, the D0 collaboration reported a large CP violation in the same-sign dimuon charge asymmetry which has the $3.2 \sigma$ deviation from the value estimated in the Standard Model. In this paper, several new physics models are considered: the MSSM, two Higgs doublet model, the recent dodeca model, and a new $Z'$ model. Generally, it is hard to achieve such a large CP violation consistently with other experimental constraints. We find that a scheme with extra non-anomalous U(1)$'$ gauge symmetry is barely consistent. In general, the extra $Z'$ gauge boson induces the flavor changing neutral current interactions at tree level, which is the basic reason allowing a large new physics CP violation. To preserve the U(1)$'$ symmetry at high energy, SU(2)$_L$ singlet exotic heavy quarks of mass above 1 TeV and the Standard Model gauge singlet scalars are introduced.
\end{abstract}

\pacs{11.30.Hv,11.30.Er,12.15.Ff}

\keywords{Dimuon asymmetry, B meson mixing, CP violation, Z prime}
\maketitle

\section{Introduction}
\label{sec:Introduction}

Discrete symmetries related to the charge conjugation (C) and parity (P) operations are of fundamental importance in weak and strong interactions. The strong interaction seems to preserve the CP symmetry at very high level of accuracy \cite{KimRMP10} while the weak interaction violates it substantially \cite{BrancoCP99}. So far, the Cabibbo-Kobayashi-Maskawa (CKM) mechanism \cite{CKM73} for the weak CP violation through quark mixing has been very successful in explaining the CP violation observed in the neutral $K$, $D$, and $B$ meson decays \cite{PData08}.

Recently, however, the D0 collaboration reported the observation of a large CP violation in the same-sign dimuon charge asymmetry from the $B$ meson decays with the $6.1 \text{fb}^{-1}$ of data \cite{Abazov:2010hv}. The same-sign dimuon asymmetry from the semi-leptonic ($s\ell$) decay of $B_{s,d}$ meson is given by,
\dis{
A_{s\ell}^b=\frac{N^{++}-N^{--}}{N^{++}+N^{--}},
}
where $N^{++}$ corresponds to each $B$ hadron decaying semileptonically to $\mu^+ X$, and similarly $N^{--}$ to $\mu^- X$. The D0 collaboration observed the asymmetry of
\dis{
A_{s\ell}^b = - (9.57 \pm 2.51 \pm 1.46) \times 10^{-3} ,\label{eq:D0exp}
}
which is about a $3.2 \sigma$ deviation from the value predicted in the Standard Model (SM) of $(-3.10^{+0.83}_{-0.98}) \times 10^{-4}$ \cite{Lenz:2010gu}. To explain this, therefore, an additional CP violation source(s) is strongly required in the $B_{s,d}$ mixing.

The D0 result has attracted a great deal of attention by introducing new TeV scale particles, such as the leptoquark models \cite{Dighe:2010nj}, the MSSM with
non-minimal flavor violation \cite{nmfvsusy}, an $R$-parity violating supersymmetric model \cite{Deshpande:2010hy}, a split SUSY model \cite{Endo:2010fk}, $Z'$ models \cite{Deshpande:2010hy,Alok:2010ij}, a fourth generation model \cite{Eberhardt:2010bm}, additional right-handed currents \cite{Alok:2010ij,rightw}, an additional SM SU(2)$_L$ doublet scalar not developing a vacuum expectation value (VEV) \cite{Trott:2010iz}, and an extra SUSY Higgs doublet \cite{Kubo:2010mh}, etc \cite{D0Interests}.
Actually, before the D0 charge asymmetry report, the direction of the new physics (NP) model has been focussed on the suppression of the additional CP violating or flavor changing neutral current (FCNC) source \cite{Gabbiani:1996hi}. Namely, the suppression of the NP CP violation was a prime interest \cite{Choi:1993yd}. After the D0 report, however, the trend has been changed to obtain a sizable NP contribution.

In this paper, we analyze the MSSM, a two Higgs doublet model, a dodeca model, and a $Z'$ model. To obtain the large enough enhancement in the same-sign dimuon asymmetry within $1 \sigma$ limit of (\ref{eq:D0exp}), the NP contribution to $\Gamma_{12}$ is limited to a value comparable to the SM contribution.
We will show that a $Z'$ model(s) has a parameter space allowing this enhancement. For the theoretically viable $Z'$ model, we consider the case that the U(1)$'$ quark charges are assigned to be flavor non-universal, which is in fact a general phenomenon with an additional $Z'$.
To preserve the extra non-anomalous U(1)$'$ symmetry at high energy, SU(2)$_L$ singlet exotic heavy quarks of mass above 1 TeV and the SM gauge singlet scalars are introduced. We also consider the other experimental results to constrain the parameter space.

This paper is organized as follows.  In Sec. \ref{sec:setup}, the notation for the $B_{s,d}$ mixing is presented. In Sec. \ref{sec:NewPhys},  we consider the various NP models such as the MSSM, two Higgs doublet model, and the dodeca model. Then, the flavor non-universal $Z'$ model is presented in Sec. \ref{sec:zprime} with an analysis on the constraints from several experiments. Sec. \ref{sec:conclusions} is a conclusion.


\section{The $B_{s,d}$ mixing}
\label{sec:setup}

The $B_q - \overline{B}_q$ oscillations for $q=s,d$ are described by a Schr\"odinger equation
\begin{eqnarray}
i \frac{\text{d}}{\text{d} t} \left( \begin{array}{c} \vert B^0 \rangle \\
  \vert \overline{B}^{\,0}  \rangle
  \end{array}  \right) = \left( M - i \frac{\Gamma}{2}\right) \left( \begin{array}{c} \vert B^0 \rangle \\
  \vert \overline{B}^{\,0} \rangle
  \end{array}  \right) ,
\end{eqnarray}
where $M$ and $\Gamma$ are the $2 \times 2$ Hermitian  mass and decay matrices. The differences of masses and widths of the physical eigenstates are given by the off-diagonal elements by \cite{Lenz07}
\begin{eqnarray}
\Delta M_q = 2 |M_{12}^q| ~, \hspace{0.5cm} \Delta \Gamma_q = 2 |\Gamma_{12}^q| \cos\phi_q~,
\end{eqnarray}
up to numerically irrelevant corrections of order $m_b^2 / M_W^2$. The CP phase difference between these quantities is defined as
\begin{eqnarray}
\phi_q = \mbox{Arg.}\left(-\frac{M_{12}^q}{\Gamma_{12}^q}\right) ,
\end{eqnarray}
where the SM contribution to this angle is
\begin{eqnarray}
\phi_d^{\text{SM}} = (-9.6^{+4.4}_{-5.8}) \times 10^{-2}~,\ \phi_s^{\text{SM}} = ( 4.7^{+3.5}_{-3.1}) \times 10^{-3} .
\end{eqnarray}

The {\em wrong} sign charge asymmetries appear in the semileptonic $B_d$ and $B_s$ decays as
\dis{
&a_{s\ell}^d \equiv \frac{\Gamma(\overline{B}_d \to \mu^+ X) - \Gamma(B_d \to \mu^- X)}{\Gamma(\overline{B}_d \to \mu^+ X) + \Gamma(B_d \to \mu^- X)}~, \\
&a_{s\ell}^s \equiv \frac{\Gamma(\overline{B}_s \to \mu^+ X) - \Gamma(B_s \to \mu^- X)}{\Gamma(\overline{B}_s \to \mu^+ X) + \Gamma(B_s \to \mu^- X)} ,
}
where the relation with $A_{s\ell}^b$ at 1.96 TeV is
\dis{
A_{s\ell}^b = (0.506\pm 0.043) a_{s\ell}^d + (0.494\pm 0.043) a_{s\ell}^s ~,
\label{eq:Aarelation}
}
which leads to roughly the $50\%:50\%$ production of the same sign dileptons from the  $b\bar d(d\bar b)$ and $b\bar s(s\bar b)$ mesons.
Considering the current experimental value of $a_{s\ell}^d = -(4.7 \pm 4.6) \times 10^{-3}$ from the B factories \cite{Barberio:2008fa}, the value of $a_{s\ell}^s$ is then obtained as \cite{Abazov:2010hv}
\begin{eqnarray}
a_{s\ell}^s = - (14.6 \pm 7.5) \times 10^{-3} . \label{eq:asls}
\end{eqnarray}
We may use the average value of $(a_{s\ell}^s)_{\text{ave}} = -(12.7 \pm 5.0) \times 10^{-3}$ considering the previous CDF result of 1.6 fb$^{-1}$ and the direct D0 measurement of the flavor specific asymmetry \cite{Abazov:2009wg}, which has about $2.5 \sigma$ deviation from the SM prediction $a_{s\ell}^{s \rm SM} = (2.1 \pm 0.6) \times 10^{-5}$ \cite{Lenz07}.

The wrong sign charge asymmetry $a_{s\ell}^q$ is related to the mass and width differences ($M_{12}$ and $\Gamma_{12}$) in the $B_q - \overline{B}_q$ system as
\begin{equation}
  a_{s\ell}^q =\mbox{Im} \frac{\Gamma_{12}^q}{M_{12}^q} = \frac{|\Gamma_{12}^q|}{|M_{12}^q|} \sin \phi_q = \frac{\Delta \Gamma_q}{\Delta M_q} \tan \phi_q\,. \label{eq:asanal}
\end{equation}
The experimental value of $\Delta M_s$ is obtained by the combination of the CDF measurement such that \cite{Barberio:2008fa}
\dis{
\Delta M_s &=17.77\pm 0.10(\rm stat.)\pm 0.07(\rm sys.)~ {\rm ps}^{-1}\\
&=(11.7\pm 0.07\pm 0.05)\times 10^{-12} ~\gev . \label{eq:DMexp}
}
The combined result of CDF and D0 is $\Delta M_s =17.78\pm 0.12~
{\rm ps}^{-1}$. With $\Gamma^s_{12} = \Gamma^{s,\, \text{SM}}_{12}$ only, however, it is impossible to obtain the observed average value $(a_{s\ell}^s)_{\text{ave}}$ from Eqs. (\ref{eq:asanal}) and (\ref{eq:DMexp}) for $q=s$ even we assume $\sin\phi_s = 1$. Therefore, an additional NP contribution to $\Gamma_{12}^s$ is preferred \cite{Dighe:2010nj, Dobrescu:2010rh,Bauer:2010dga}. This feature triggered the recent NP approaches, which is different from the old approach trying to confirm the SM at the electroweak scale \cite{Hagelin81,Buras98,Cahn99}.

\begin{figure}
\begin{center}
\epsfig{figure=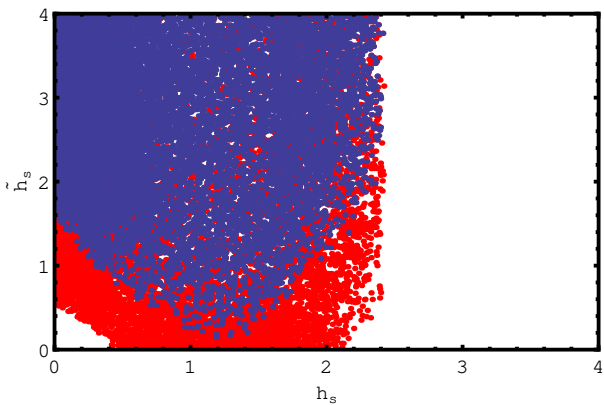,width=8cm}
\end{center}
\caption{The parameter space on $\tilde{h}_s$ versus $h_s$. The points  satisfy Eq. (\ref{eq:DMexp}) and the averaged value of $(a_{s\ell}^s)_{\text{ave}}$. The blue points are within $1 \sigma$, while the red points within $2 \sigma$. The deviation from the SM value, $h_s = \tilde{h}_s = 0$, is about $\sim 2.5 \sigma$.}
\label{fig:hsratio}
\end{figure}

\begin{figure}
\begin{center}
\epsfig{figure=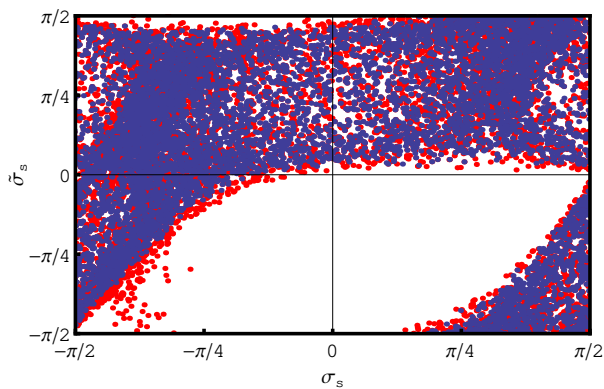,width=8cm}
\end{center}
\caption{The parameter space on $\tilde{\sigma}_s$ versus $\sigma_s$. The points are described in Fig. \ref{fig:hsratio}. The deviation from the SM value, ${\sigma}_s = \tilde{\sigma}_s = 0$, is about $\sim 2.5 \sigma$.}
\label{fig:sigsratio}
\end{figure}

To probe the NP contribution, we split $\Gamma_{12}$ or $M_{12}$ to the SM and NP contributions as
\begin{eqnarray}
\frac{\Gamma_{12}^{q\, \rm NP}}{\Gamma_{12}^{q\, \rm SM}} \equiv \tilde{h}_q e^{i 2 \tilde{\sigma}_q}~,~
\frac{M_{12}^{q\, \rm NP}}{M_{12}^{q\, \rm SM}} \equiv h_q e^{i 2\sigma_q}~,
\end{eqnarray}
for real and non-negative parameters $\tilde{h}_q$ and $h_q$, with the phases constrained in the region, $0 \le \sigma_q, \tilde{\sigma}_q \le \pi$. From Eq. (\ref{eq:asanal}), then the wrong sign charge asymmetry is given by
\begin{eqnarray}
a_{s\ell}^q = \mbox{Im} \frac{\Gamma_{12}^q}{M_{12}^q} = \mbox{Im} \left(\frac{\Gamma_{12}^{q}}{\Gamma_{12}^{q\, \rm SM}} \cdot \frac{\Gamma_{12}^{q\, \rm SM}}{M_{12}^{q\, \rm SM}} \cdot \frac{M_{12}^{q\, \rm SM}}{M_{12}^q} \right)~,
\end{eqnarray}
which is expressed as
\begin{widetext}

\dis{ a_{s\ell}^q &= \frac{|\Gamma_{12}^{q\, \rm SM}|}{|M_{12}^{q\, \rm SM}|} \frac{-1}{1+h_q^2+2h_q\cos2\sigma_q} \left\{\tilde{h}_q \sin2\tilde{\sigma}_q (1+h_q\cos2\sigma_q)- h_q \sin2\sigma_q (1+\tilde{h}_q\cos2\tilde{\sigma}_q)\right\}~,
}
where $\phi_q^{\rm SM} \ll \pi$ are applied.
\end{widetext}
With this relation, we can explore the possible parameter space in the following figures for $q=s$, satisfying the average value on $a_{s\ell}^s$. Inserting the central value of the SM prediction, $|\Gamma_{12}^{s\, \rm SM} / M_{12}^{s\, \rm SM}| = (4.97 \pm 0.94) \times 10^{-3}$ \cite{Lenz07}, from Eq. (\ref{eq:DMexp}) we obtain the relations on $\tilde{h}_s - h_s$ in Fig. \ref{fig:hsratio} and on $\tilde{\sigma}_s - \sigma_s$ in Fig. \ref{fig:sigsratio}. The blue points satisfy the value of $(a_{s\ell}^s)_{\text{ave}}$ with the $1 \sigma$ limit while the red points with $2 \sigma$ limit. In both figures, we can clearly see the deviation from the SM value, $h_s = \tilde{h_s} = \sigma_s = \tilde{\sigma}_s = 0$, by about $2.5 \sigma$. It seems that about $\mathcal{O}(1)$ $\tilde{h}_s$ is generically required to obtain the averaged observed value of $(a_{s\ell}^s)_{\text{ave}}$ within $1 \sigma$.

\section{New physics for the additional CP violation sources}
\label{sec:NewPhys}

In the SM, the width difference $\Gamma_{12}^{s\, \rm SM}$ is dominantly given by the CKM matrix, $b\to c+(\bar cs)$. Therefore, the NP operators in the electroweak Hamiltonian which provide $\mathcal{O}(1)$ $\tilde{h}_s$ must have the Wilson coefficient of order $V_{cb} \sim \lambda^2$, where $\lambda \sim 0.2$ is the Cabibbo angle . However, such operators are highly constrained by the various experimental limits on B meson decays. There are only two possible dimension six operators of the SM fermions for the viable NP contribution to $\Gamma_{12}^s$. They are $(\bar{b}s)(\bar{\tau}\tau)_{V,A}$ and $(\bar{b}s)_{V+A} (\bar{c}c)_{V-A}$ as analyzed in \cite{Bauer:2010dga}.
The NP contribution to the total lifetime $\tau_{B_s}$ due to these operators is about $10 \%$ order. This does not contradict the observed result containing about $2 \%$ error bar since we have already taken into account a large theoretical uncertainty when we calculated the SM value. On the other hand, the ratio $\tau_{B_s}/\tau_{B_d}$ is more precisely predicted since the theoretical uncertainties due to unknown nonperturbative effects are canceled in the ratio, and hence there is not a much space for a NP contribution to this quantity. Therefore, if the ratio of the $B_s$ and $B_d$ lifetime precisely agrees with the SM value by $1+\mathcal{O}(1 \%)$, the additional contribution in the $B_d$ system is required to compensate the $10 \%$ NP contribution to $\tau_{B_s}$.\footnote{The operator $\bar{b}d\bar{u}c$ is the only allowed NP contribution to the $B_d$ system, which is not constrained by other experiments \cite{Bauer:2010dga}.}
\begin{figure}[t!]
 \begin{center}
 \epsfig{figure=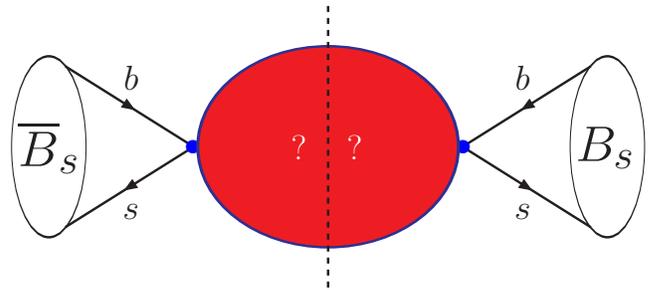,width=8.5cm,height=3.8cm,angle=0}
\caption{The effective $B_s-\overline{B}_s$ mixing diagram. The four fermion interaction is the square of bilinears $\sum_{A,B}(\overline{s}\,\Gamma^Ab)(1/M^2)(\overline{s}\,\Gamma^Bb) $ where the blue bullets are $\Gamma^{A,B}$. The red oval appears as an effective suppression mass squared.}
\label{fig1:EffBs}
\end{center}
\end{figure}

For the neutral $B_s$ system, the $\Delta B=2$ and $\Delta S=2$ diagrams are schematically shown in Fig. \ref{fig1:EffBs}. The four fermion interaction is represented as the square of bilinears $(\overline{s}\,\Gamma^Ab)(1/M^2)(\overline{s}\,\Gamma^Bb) $, and in Fig. \ref{fig1:EffBs} the blue bullets represent the Dirac matrices $\Gamma^{A,B}$. The cut diagram by the dashed line gives the absorptive part. The red oval appears as an effective suppression mass in the four fermion interaction. Suppressing the Dirac $\Gamma$ matrices and color indices, the bilinears are
\dis{
\overline{s}_L b_L,\quad \overline{s}_R b_R,\\
 \overline{s}_L b_R,\quad \overline{s}_R b_L,
}
where the first line contains $\gamma_\mu$ or $\gamma_\mu\gamma_5$, and the second line contains ${\bf 1}, \gamma_5, \sigma_{\mu\nu}$ or
$\sigma_{\mu\nu}\gamma_5$. All these are summarized as five operators in Ref. \cite{UTfit08}. The mass difference $M_{12}^s$ is obtained from the dispersive part of the mixing diagram Fig. \ref{fig1:EffBs}. The width difference $\Gamma_{12}^s$ is obtained from the absorptive part of Fig. \ref{fig1:EffBs}, which arises from the $B_s$ decays to final states of zero strangeness, indicated by the dashed lines. In this section, we analyze the MSSM, a two Higgs doublet model, and a dodeca model to search for possibilities of obtaining a large same-sign dimuon asymmetry.

\subsubsection{The MSSM}

\begin{figure}[t!]
 \begin{center}
 \epsfig{figure=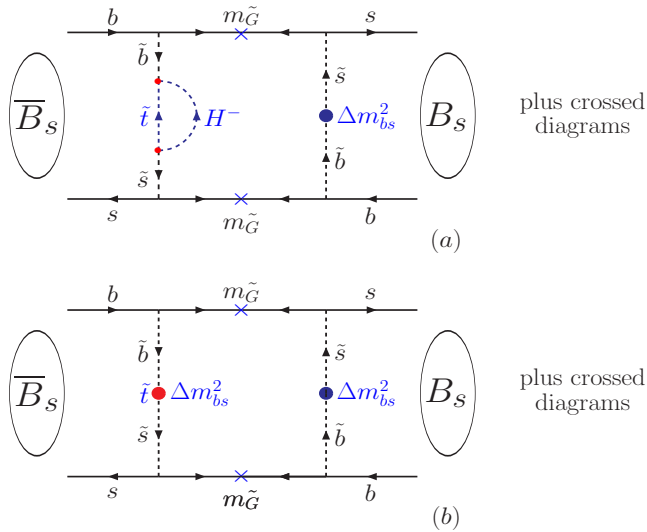,width=8.5cm,height=7cm,angle=0}
\caption{The $B_s-\overline{B}_s$ mixing through the gluino mass:  ($a$) A detail of one mass mixing and  ($b$) all mass mixings without details. The diagram with the charged gauginos which is a mere supersymmetrization of the SM FCNC is also possible, but with smaller gauge couplings. ($a$) is drawn again in ($b$). The red bullet in ($b$) contains a CP phase whose origin is shown in ($a$). The $A$ terms are colored red, and the box diagram of ($a$) has an unremovable CP phase.}
\label{fig1:Loop}
\end{center}
\end{figure}

Supersymmetry (SUSY) is one of the most promising candidates of the NP beyond the SM. The minimal SUSY models \cite{Gabbiani:1996hi} where the $R$-parity is preserved and the K\"ahler metric is flat are constructed not to provide the additional large CP-violating or FCNC sources.
These SUSY models are the low energy limit of the spontaneously broken minimal $N=1$ supergravity theories with flat K\"ahler metric. This minimal case with soft SUSY breaking insertions is depicted in Fig. \ref{fig1:Loop}.
The gluino mass may introduce a phase, however, we need an interference with the $A$ and/or $B$ terms with the gluino phase. In Fig. \ref{fig1:Loop}, the red bullet in ($b$) contains a CP phase whose origin is shown in ($a$). The $A$ terms are colored red, and the box diagram of ($a$) has an unremovable CP phase. Because both the gluino mass and the $A$ terms appear in Fig. \ref{fig1:Loop}($a$), this box diagram has an unremovable CP phase. $M_{12}^{s\, \rm NP}$ can be obtained from Fig. \ref{fig1:Loop}, however, it is quite suppressed since it is basically a loop diagram. In addition, there is no $\Gamma_{12}^{s\, \rm NP}$ with new particles heavier than $m_{B_s}$. Therefore, it is impossible to provide the observed central value of $a_{s\ell}$ in the minimal case.

In the non-minimal flavor violation case which is originally obtained from assuming a non-flat K\"ahler metric, it may be possible to enhance the same-sign dimuon asymmetry more. In this case, the origin of the red bullet in Fig. \ref{fig1:Loop} is not the loop effect. However, it is still about $1 \sigma$ away from the central value of $(a_{s\ell}^s)_{\text{ave}}$ and highly constrained by $B \to X_s \gamma$ result with moderate $\tan\beta \lesssim 10$ so that new approach to the family symmetry is required to obtain the central value \cite{nmfvsusy}.

\subsubsection{Two Higgs-doublet model without SUSY}

If we do not consider SUSY, we have to look for appropriate scalar or gauge boson which mediates fermion currents. In this case, the magnitude and phase of coupling or mass of intermediate boson impose stringent bound for the same-sign dimuon asymmetry. We analyze the following interaction with two Higgs doublets $H_u$ and $H_d$,
\dis{
{\cal L}_Y &=f^{(u)}_{ij}\overline{Q}_L^{\,i} u_R^j H_u
+f^{(d)}_{ij}\overline{Q}_L^{\,i} d_R^j H_d\,,\\
&Y(H_u)=-\frac12,~ Y(H_d)=+\frac12\,.
}
The CKM mixing matrix is given by $V_{\rm CKM} = V_u V_{d}^{\dagger}$. The interaction between the quarks and the charged Higgs are given in terms of the mass eigenstates by \cite{Abbott:1979dt}
\begin{widetext}
\dis{
  \frac{g}{2 \sqrt{2} M_W}& \Big\{H_2^+ \overline{U}\left(\frac{1}{\tan \beta} M_u V_{CKM}(1+\gamma_5) + V_{CKM} M_d \tan \beta (1 -\gamma_5) \right) D \\
+&  H_2^- \overline{D} \left(\frac{1}{\tan \beta}V^{\dagger}_{CKM} M_u (1-\gamma_5) +
  \tan \beta M_d V^{\dagger}_{CKM} (1+ \gamma_5)  \right)U \Big\}~,
}
\end{widetext}
where $M_u=\mathrm{diag}(m_u,~ m_c,~ m_t)$ and $M_d= \mathrm{diag}(m_d,~ m_s,~ m_b)$. Interactions between different chiral states mimic the charged weak current couplings but with the coefficients $\tan\beta$ or $1/\tan\beta$.

\begin{figure}
 \begin{center}
 \epsfig{figure=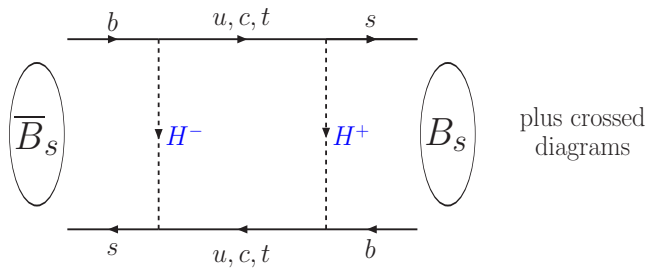,width=8.5cm,height=3.5cm,angle=0}
\caption{The $B_s-\overline{B}_s$ mixing through the charged Higgs. The $W$ line can be replaced with the charged Higgs.}
\label{fig2:TwoHiggs}
\end{center}
\end{figure}

\begin{figure}
 \begin{center}
 \epsfig{figure=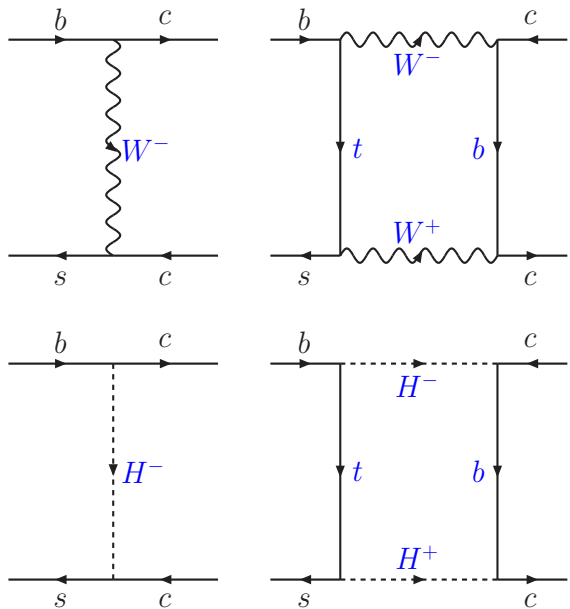,width=7.5cm,height=8cm,angle=0}
\caption{The $\overline{B}_s$ decay diagrams. The tree level decays and loop effects introducing a CP phase.}
\label{fig2:TwoHiggsCP}
\end{center}
\end{figure}

Figure \ref{fig2:TwoHiggs} is relevant for the calculation of $M_{12}$ and $\Gamma_{12}$. The $\overline{B}_s$ decay diagrams are depicted in Fig. \ref{fig2:TwoHiggsCP}. Intermediate charged Higgs may be replaced by $W$ boson. We are interested in $\Gamma_{12}$ to see whether it can introduce a
large enhancement. For that we take the interference terms from Fig. \ref{fig2:TwoHiggsCP}. The top two diagrams leads to the SM prediction which is known to be small. The bottom two diagrams are the NP contribution. The interference does not appear between top and bottom diagrams due to the chirality difference. To achieve the central value of $(a_{s\ell}^s)_{\rm ave}$, the left two diagrams must be of the same order. The CP phase from the NP sector is not enhanced since it is the same as that by the CKM elements. In addition, the loop factor in the two Higgs doublet model is not easily enhanced compared to the SM one since coupling to charged Higgs is of the order $(\mathrm{quark~ mass})/(M_W)$, which is Yukawa suppressed.
To obtain the enhancement by $\Gamma_{12}^{s\, \rm NP}$ comparable to $\Gamma_{12}^{s\, \rm SM}$, a large $\tan\beta \gtrsim 170$ or small $\tan\beta$ $\lesssim 0.03$ are required. However, this parameter region is outside the perturbative region the Yukawa couplings at the electroweak scale $0.3 \lesssim \tan\beta \lesssim 120$ \cite{Haber:1997dt,Barger:1989fj}. (Assuming that the low energy theory at the electroweak scale is the MSSM, and there is no additional new physics below the GUT scale, the Yukawa couplings remain finite at all energy scales below the GUT scale if $1.5 \lesssim \tan\beta \lesssim 65$ for $m_t = 170$ GeV \cite{Haber:1997dt}.) Therefore, a non-conventional approach similar to the uplifted Higgs model \cite{Dobrescu:2010rh} is required to obtain the central value of $(a_{s\ell}^s)_{\rm ave}$. On the other hand, we may consider the enhancement of the asymmetry which is more than $1 \sigma$ away from the observed central value with smaller $\tan\beta < 170$.\footnote{We may enhance $M_{12}$ by the top quark exchange in Fig. \ref{fig2:TwoHiggs}} In such scenarios, the experimental constraints such as $B_s \to \mu^- \mu^+$, $B \to \tau \nu$, and $B \to D \tau \nu$ must be considered like the scenarios in the MSSM \cite{largetansusy}.

\subsubsection{A dodeca-model}

Recently, two of us introduced a {\em dodeca} model to explain the mixing angles of the lepton and quark sectors using a discrete symmetry $D_{12}$\cite{KimSeo1}, which introduces multiple Higgs particles. On the other hand, the Higgs fields are assigned as singlets of the discrete group while the needed $D_{12}$ transformation property in the Yukawa couplings is provided by the flavons which are the SM singlets decoupled at high energy. In this case, there will be no distinction from the Kobayashi-Maskawa model prediction. Therefore, to have any change from the SM, we consider the original multi Higgs model \cite{KimSeo1}. Even in this case, however, the coupling of multi Higgs to quark current should be Yukawa suppressed, and as a result a large enhancement in the same-sign dimuon asymmetry is not expected. Here, we show that indeed this happens.

\begin{figure}
 \begin{center}
 \epsfig{figure=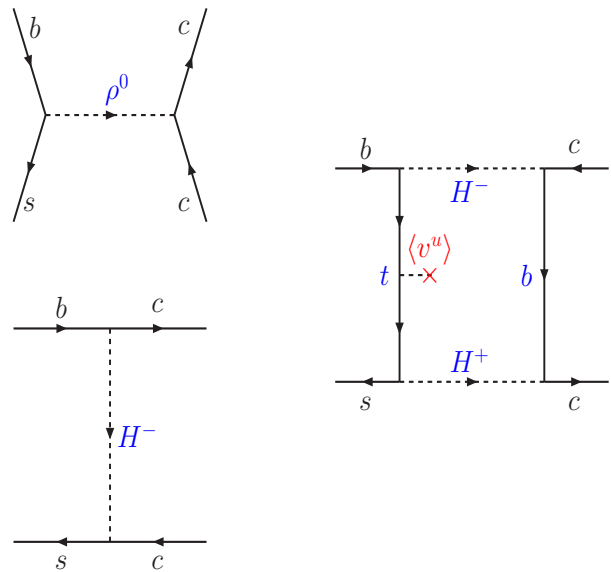,width=8cm,height=7.5cm,angle=0}
\caption{The $\overline{B}_s$ decay diagrams in the dodeca model. The charged Higgs diagram interfere with the loop diagram. }
\label{fig6:dodecaCP}
\end{center}
\end{figure}

In this dodeca model, there are two ways to introduce the NP contributions, one by the charged Higgs and the other by the neutral Higgs.\footnote{For the two Higgs doublet model, we avoided the FCNC by coupling only one Higgs doublet to the same charge quarks.}
 For the charged Higgs boson exchange, the idea is similar to the two Higgs doublet model except the fact that the couplings are not proportional to the CKM coefficients. The CP violation is introduced by the loop with charged Higgs bosons in the $\overline{B}_s$ decay process, as shown in Fig. \ref{fig6:dodecaCP}. The NP CP violation is introduced by the interference of two types of trees and one type of the loops. There is no loop diagram through the neutral Higgs exchange.
The contribution to $\Gamma_{12}^{s\, \rm NP}$ is obtained as
\dis{& \Gamma_{12}^{s\, \rm NP} \sim
 \frac{\vert f^{(d)I}_{22} f^{(d)I}_{23} \vert ^2}{M^4_{\rho_I^{(d)+}}}
 M_{B^0_s}^5
  \\
  & =\frac{m_2^2 m_3^2}{8 v^4 M^4_{\rho_I^{(d)+}}} \epsilon^2 M_{B^0_s}^5
  = 3.76 \cdot 10^{-11}(\mathrm{GeV})^5 \times \frac{1}{M^4_{\rho_I^{(d)+}}}~,
   }
where the parameters are defined in Appendix \ref{appendix:dodeca}.
Since  $M^4_{\rho_I^{(d)+}}$ is expected to be larger than $(\rm 100~ GeV)^4 $ order, this indicates that the NP contribution to $\Gamma_{12}$ is not enough to enhance the dimuon asymmetry to the observed central value, due to the Yukawa suppression.


\section{A $Z^{\prime}$ model}
\label{sec:zprime}

We can consider the models with the extra U(1)$^{\prime}$ gauge bosons with flavor non-universal couplings to the SM weak gauge eigenstates.
The existence of an extra U(1) gauge symmetry \cite{uone} is motivated from the grand unified theories (GUTs), superstring models, and models with large extra dimensions \cite{Pomarol99}. In perturbative heterotic string models with supergravity mediated SUSY breaking, the mass of $Z^{\prime}$ is expected to be less than around a TeV, which is naturally induced when the U(1)$^{\prime}$ breaking is driven by a radiative mechanism \cite{stringzprime}.\footnote{
In this case the two Higgs doublets can also have the U(1)$^{\prime}$ charges to forbid the exact $\mu$ term, while allowing the effective $\mu$ and $B\mu$ terms to be generated at the U(1)$^{\prime}$ scale.}
The neutral current Lagrangian with one extra neutral gauge boson  $Z_{2, \mu}^0$, with an additional Abelian gauge symmetry U(1)$'$, can be written as \cite{Langacker:2000ju}
\begin{eqnarray}
\mathcal{L}_{\text{NC}} = -e J^{\mu}_{\text{\small em}} A_{\mu} - g_1 J^{(1) \mu} Z_{1, \mu}^0 - g'' J^{(2) \mu} Z_{2, \mu}^0 ,
\end{eqnarray}
where $Z_1^0$ is the SU(2)$_L\times$U(1) neutral gauge boson $Z$ if there is no mixing with $Z_2^0$. The U(1) coupling is $g_1 = g/\cos\theta_W$ where $g$ is the coupling constant of SU(2)$_L$. The currents are given as
\dis{
J_{\mu}^{(1)} &= \sum_i \bar{\psi}_i \gamma_{\mu} \left[ \epsilon_L (i) P_L + \epsilon_R (i) P_R \right] \psi_i , \\
J_{\mu}^{(2)} &= \sum_{i,j} \bar{\psi}_i \gamma_{\mu} \left[ \epsilon_{\psi_{L_{ij}}}^{(2)} P_L + \epsilon_{\psi_{R_{ij}}}^{(2)} P_R \right] \psi_j ,
}
where the sum is over all quarks and leptons $\psi_{i,j}$ and $P_{L,R}=(1 \mp \gamma_5 )/2$. The values of $\epsilon_{\psi_{L_{ij}}}^{(2)}$ and $\epsilon_{\psi_{R_{ij}}}^{(2)}$ are the chiral couplings of the new gauge boson, which are real. The SM chiral couplings are \cite{Kim81}
\dis{
\epsilon_R (i) = - \sin^2\theta_W Q_i , \hspace{0.2cm} \epsilon_L (i) = t_3^i -\sin^2\theta_W Q_i ,
}
where $t_3^i$ and $Q_i$ are the third component of the weak isospin and the electric charge of the fermion $i$, respectively. $\theta_W$ is the weak mixing angle.

When $\epsilon^{(2)}$ are non-diagonal, we can obtain the flavor changing effects immediately. There have been works on this possibility, which have the exotic fermions mixing with the SM fermions \cite{nondiagonalzp} as in the E$_6$ models. If the $Z_2$ couplings are diagonal and non-universal, flavor changing effects arise by non-zero fermion mixing \cite{Barger:2004qc,Cheung:2006tm,He:2006bk}. Here, we consider the case that $\epsilon^{(2)}$ of the left handed quarks are not flavor-universal. The fermion Yukawa matrices are diagonalized by the unitary matrices $V_{L, R}^{\psi}$ where the CKM matrix is
\begin{eqnarray}
V_{\text{CKM}} = V_L^{u} V_L^{d\dagger} .
\end{eqnarray}
Therefore, the chiral $Z_2^0$ couplings in the fermion mass eigenstate basis are
\dis{
B_{ij}^{\psi_L} \equiv ( V_L^{\psi} \epsilon_L^{(2)} V_L^{\psi^{\dagger}} )_{ij} , \hspace{0.2cm} B_{ij}^{\psi_R} \equiv ( V_R^{\psi} \epsilon_R^{(2)} V_R^{\psi^{\dagger}} )_{ij}\, .
}
With a diagonal $\epsilon_R^{(2)}$, the right-handed sector remains flavor-diagonal.

Now, for a convenience sake let us suppose that $V_{L}^{u}=V_{R}^{u}=V_{R}^{d}=I$ and $V_{L}^{d}=V_{CKM}^{\dagger}$. Then, with the Wolfenstein parametrization \cite{Wolfenstein83},
\begin{widetext}
\dis{ V_{CKM} =
  & \left ( \begin{array}{ccc}
   1-\frac{\lambda^2}{2} & \lambda & A \lambda^3 (\rho - i \eta + \frac{i}{2} \eta \lambda^2)  \\
   -\lambda & 1- \frac{\lambda}{2} - i \eta A^2 \lambda^4 & A \lambda^2 ( 1+i \eta \lambda^2 )\\
    A \lambda^3 (1- \rho - i \eta) & -A \lambda^2 & 1
  \end{array} \right)
}
\end{widetext}
where the imaginary part is written up to order $\lambda^5$ and the real part up to order $\lambda^3$.  For simplicity, let us represent the U(1)$'$ operator for our left-handed quarks, \ie $\epsilon_{u_L,d_L}^{(2)}$, as a diagonal but flavor non-universal matrix such that for a possibility of $bs$ flavor changing coupling with $x \ne 1$
\dis{ \epsilon_{u_L,d_L}^{(2)} =
  & C_{q_L}\left ( \begin{array}{ccc}
   x & 0 & 0  \\
   0 & x & 0 \\
  0 & 0 & 1
  \end{array} \right),\label{eq:qnonuniv}
}
while $\epsilon_{u_R, d_R}^{(2)}$ are assumed to be diagonal and flavor universal such that
\dis{ \epsilon_{u_R, d_R}^{(2)} =
  & C_{u_R, d_R}\left ( \begin{array}{ccc}
   1 & 0 & 0  \\
   0 & 1 & 0 \\
  0 & 0 & 1
  \end{array} \right).
}
Then, the U(1)$'$ couplings to the mass eigenstates of left-handed down type quarks are given approximately by
\begin{widetext}
\dis{
   B^{d_L}= V_{CKM}^{\dagger} \epsilon_{d_L}^{(2)} V_{CKM} \simeq C_{q_L}\left ( \begin{array}{ccc} x  & (x-1) i A^2 \lambda^5 \eta & -(x-1) A \lambda^3 (1-\rho+i \eta)  \\
   -(x-1) i A^2 \lambda^5 & x & (x-1)A \lambda^2 + i x A \lambda^4 \eta \\
  -(x-1) A \lambda^3 (1-\rho-i \eta) & (x-1)A \lambda^2 - i x A \lambda^4 \eta & 1
  \end{array} \right)~,
}
\end{widetext}
and the same for the left-handed up type quarks also. Thus, $B^{d_L}_{12}$ which contributes to the neutral K meson system is of order $\lambda^5$, while $B^{d_L}_{23}$ which is relevant for the current neutral $B_s$ meson system is of order $\lambda^2$. So, it is possible to raise the $B_s$ decay CP phase by a factor of $\lambda^{-3}\sim 10^2$ without tampering the phenomenology of the neutral K meson system very much.

\begin{figure}
\begin{center}
\epsfig{figure=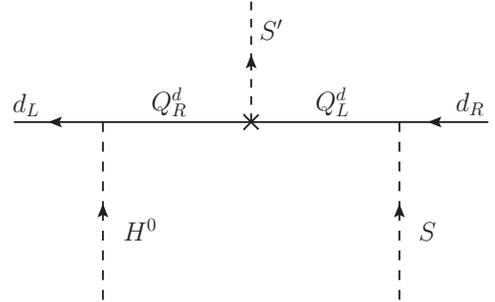,width=0.4\textwidth,angle=0}
\caption{A typical Yukawa interaction between the first and the second generation quarks. It is obtained by the insertion of the exotic SU(2)$_L$ singlet heavy quarks $Q_{L,R}^{u,d}$, SM singlet scalars $S$ and $S'$ preserving the SU(2)$ \times$U(1)$_Y \times $U(1)$^{\prime}$ symmetry.  }
\label{fig:heavyquark}
\end{center}
\end{figure}

At high energy where the extra U(1)$'$ symmetry is restored, the Yukawa interactions must preserve the symmetry. Therefore, additional SM singlet scalar field $S$ with non-zero U(1)$'$ charge must be introduced due to the flavor non-universal U(1)$'$ charge assignment to the left-handed quark doublets. We now consider that the Yukawa interactions of the light quarks are generated by the higher dimensional operators, {\it \`a la} Froggatt and Nielsen \cite{Froggatt:1978nt} but at the TeV scale,
$\bar{q}_{L}^i H q_{R}^j (S/M)$ for the quark flavors $i=1,2$ and $j=1,2,3$. The TeV scale mass $M$ can be introduced by assuming the existence of an exotic heavy quark with mass $M$, whose Yukawa couplings to the SM quarks are dictated by the U(1)$'$ symmetry. In order to preserve the U(1)$_Y$ symmetry in the Yukawa couplings at high energy, the exotic heavy quarks interacting with the down-type and up-type quarks must be different. Therefore, we define $Q_{L,R}^d$ and $Q_{L,R}^u$ for such exotic quarks which have different U(1)$_Y$ charges. Now, considering the anomaly free U(1)$'$, the Higgs has non-zero U(1)$'$ charge and so is $M$ in the effective Lagrangian. Therefore, it is required to introduce another SM singlet $S'$ whose non-zero VEV provides the value of $M$, simultaneously breaking the extra non-anomalous U(1)$'$ gauge symmetry. The mass of new gauge boson is given as $M_{Z'} = \frac12 g'' \sqrt{v_s^2 + v_{s'}^2}$ where $\langle S \rangle = v_s / \sqrt{2}$ and $\langle S' \rangle = v_{s'} / \sqrt{2}$. Then, the SU(2)$\times$U(1)$_Y\times$ U(1)$'$ invariant Yukawa couplings can be obtained, achieving the Froggatt-Nielson mechanism between the third and the first two families. Assuming the exotic quarks are SU(2)$_L$ singlets, the Froggatt-Nielson type higher dimensional operators are described as Fig. \ref{fig:heavyquark}.

The non-anomalous U(1)$'$ charges of the matter fermions, Higgs and singlet scalars are given in Table \ref{table:charge}. The anomaly cancelation conditions are
\begin{widetext}
\dis{
\hskip 0.2cm {\rm U(1)' SU(3)^2}:&~ 2(2x+1) C_{q_L}-3 (C_{d_R}+C_{u_R}) + 2 C_{S'} = 0
\\
\hskip 0.2cm {\rm U(1)' SU(2)^2}:&~ (2x+1) C_{q_L} + C_{\ell} = 0
\\
\hskip 0.2cm {\rm U(1)' U(1)_Y^2}:&~ \frac{1}{18} (2x+1) C_{q_L} - \frac13 (C_{d_R}+4C_{u_R}) + \frac12 C_{\ell} - C_{e_R} + \frac59 C_{S'} = 0
\\
\hskip 0.2cm {\rm U(1)_YU(1)^{\prime 2} }:&~ \frac13 (2x^2 +1) C_{q_L}^2 + C_{d_R}^2 - 2 C_{u_R}^2 - C_{\ell}^2 + C_{e_R}^2 -\frac13 (C_{Q_L^d}^2 - C_{Q_R^d}^2) + \frac23 (C_{Q_L^u}^2 - C_{Q_R^u}^2) = 0 \\
\hskip 0.2cm {\rm U(1)^{\prime 3}}:&~  2(2x^3 +1) C_{q_L}^3 - 3 (C_{d_R}^3 + C_{u_R}^3) + 2 C_{\ell}^3 - C_{e_R}^3  + (C_{Q_L^d}^3 - C_{Q_R^d}^3) + (C_{Q_L^u}^3 - C_{Q_R^u}^3) = 0~, \label{eq:anomaly}
}
where we used the relation from the Yukawa couplings, $C_{S'} = C_{Q_L^u} - C_{Q_R^u} = C_{Q_L^d} - C_{Q_R^d}$. In addition, we can consider the gravitational anomaly cancelation condition
\begin{eqnarray}
4 x C_{q_L} + 2 C_{q_L} -3 (C_{d_R}+C_{u_R}) + 2 C_{S'}  + 2 C_{\ell} - C_{e_R} = 0~. \label{eq:anomalygra}
\end{eqnarray}
\end{widetext}
The first constraint in (\ref{eq:anomaly}) and Eq. (\ref{eq:anomalygra}) give
\begin{eqnarray}
2 C_{\ell} - C_{e_R} = 0~,
\end{eqnarray}
which will turn out to be very useful. Although the U(1)$'$ assignments of Table \ref{table:charge} cannot avoid $x=1$ from the quadratic and cubic constraints of (\ref{eq:anomaly}), it is always possible to satisfy these constraints for $x \ne 1$ by introducing additional SM singlet leptons.

\begin{table}[htdp]
\begin{center}
\begin{tabular}{|c|c|c|}
\hline
Fields & $U(1)^{\prime}$ charge & $U(1)_Y$ charge \\
\hline
${u_L \choose d_L}$  ,  ${c_L \choose s_L}$ & $x C_{q_L}$ & 1/6 \\
\hline
${t_L \choose b_L}$ & $C_{q_L}$ & 1/6 \\
\hline
$u_R^{i=1,2,3}$ & $C_{u_R} = 2(x+1) C_{q_L}$ & 2/3 \\
\hline
$d_R^i$ & $C_{d_R} = -2x C_{q_L}$ & -1/3 \\
\hline
${\nu_L^i \choose e_L^i}$ & $C_{\ell} = -(2x+1) C_{q_L}$ & -1/2 \\
\hline
$e_R^i$ & $C_{e_R} = -2(2x+1) C_{q_L}$ & -1 \\
\hline
$Q_L^u$ & $C_{Q_L^u} = (x+3) C_{q_L}$ & 2/3 \\
\hline
$Q_R^u$ & $C_{Q_R^u} = (3x+1) C_{q_L}$ & 2/3 \\
\hline
$Q_L^d$ & $C_{Q_L^d} = -(3x-1) C_{q_L}$ & -1/3 \\
\hline
$Q_R^d$ & $C_{Q_R^d} = -(x+1) C_{q_L}$ & -1/3 \\
\hline
$H$ & $C_H = (2x+1) C_{q_L}$ & 1/2 \\
\hline
$S$ & $C_S = -(x-1) C_{q_L}$ & 0 \\
\hline
$S'$ & $C_{S'} = -2(x-1) C_{q_L}$ & 0 \\
\hline
\end{tabular}
\end{center}
\caption{The U(1)$'$ charge assignment.}
\label{table:charge}
\end{table}

If there exists a mixing between $Z_{1}^0$ and $Z_{2}^0$, then the mass eigenstates $Z$ and $Z'$ couple as
\begin{eqnarray}
\mathcal{L}_{NC}^{Z} &=& -g_1 \left[ \cos\theta J^{(1) \mu} + \frac{g''}{g_1} \sin\theta J^{(2)\mu}\right] Z_{\mu} \nonumber \\
& & \ -g_1 \left[ \frac{g''}{g_1} \cos\theta J^{(2) \mu} - \sin\theta J^{(1)\mu}\right] Z'_{\mu} ,
\end{eqnarray}
where $\theta$ is the mixing angle between $Z_1^0$ and $Z_2^0$. Since the mixing angle is quite suppressed as $|\theta| \lesssim 10^{-3}$ due to the electroweak precision measurements \cite{Erler:2009jh}, we just assume that there is no mixing and we can approximately treat $Z_1^0$ as $Z$ and $Z_2^0$ as $Z'$.

As commented in the previous section, $(\bar{b}s)(\bar{\tau}\tau)_{V,A}$ and $(\bar{b}s)_{V+A} (\bar{c}c)_{V-A}$ are the only possible dimension six operators which provide the $\mathcal{O}(1)$ $\tilde{h}_s$ without severe constraints from the other experiments \cite{Bauer:2010dga}. Therefore, new contributions to $\Gamma_{12}^{s\, \rm NP}$ via the $\tau$ or $c$-quark loops containing the $Z'$ exchange must be considered. Such enhancement of the dimuon asymmetry is first given in \cite{Alok:2010ij} although the U(1)$'$ symmetry preservation is not considered.

\begin{figure}
\begin{center}
\epsfig{figure=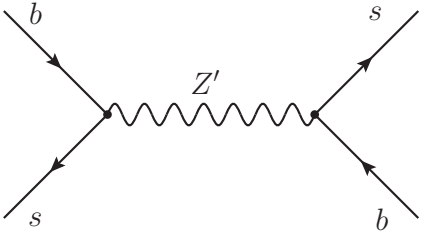,width=6cm}
\end{center}
\caption{The tree level $B_s - \overline{B}_s$ mixing via the new $Z'$ gauge boson. }
\label{fig:zprime1}
\end{figure}

\begin{figure}
\begin{center}
\epsfig{figure=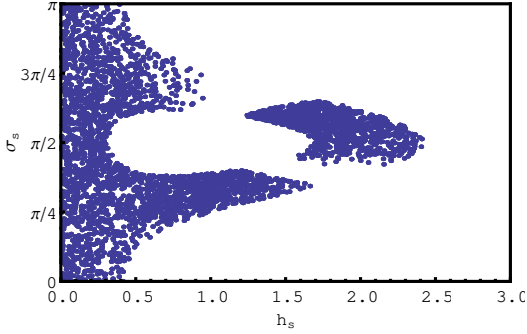,width=7cm}
\end{center}
\caption{The allowed parameter region in the $\sigma_s - h_s$ plane from the experimental bound on $\Delta M_s$ (Eq. (\ref{eq:DMexp})), $\Delta \Gamma_s$ (Eq. (\ref{eq:deltagam})), and $S_{\psi \phi}$ (Eq. (\ref{eq:spp})).}
\label{fig:hsigsbound}
\end{figure}

The mass difference $M_{12}$ is obtained from the dispersive part of the tree level mixing via $Z'$ exchange Fig. \ref{fig:zprime1} such that
\dis{
M_{12}^{\rm NP} &= \frac{4G_F}{3 \sqrt2}\left(\frac{g''}{g_1}\right)^2 \frac{M_W^2}{M_{Z'}^2 \cos^2 \theta_W} R^{6/23} (B_{sb}^{d_L})^2 f_{B_s}^2 m_{B_s} B_{B_s} \\
&=\frac{4G_F}{3 \sqrt2}\left(\frac{g''}{g_1}\right)^2 \frac{M_W^2}{M_{Z'}^2 \cos^2 \theta_W} R^{6/23} A^2 f_{B_s}^2 m_{B_s} B_{B_s}\\
& \hspace{1.5cm} \cdot C_{q_L}^2 [(x-1)^2 \lambda^4 + i 2x(x-1) \eta\lambda^6] }
where $R \equiv \alpha_s (M_W)/\alpha_s (m_b)$ and we neglected the $\mathcal{O}(\lambda^8)$ terms. The next leading order QCD corrections are not present due to our initial assumption $B_{sb}^{d_R}=0$. This result contains only one power of $G_F$ due to the tree level mixing, while the SM model result is
 \dis{M_{12}^{\rm SM} &= \frac{G_F^2}{12 \pi^2} M_W^2  (V_{tb} V_{ts}^{\ast})^2 \eta_{2B} f_{B_s}^2 m_{B_s} B_{B_s} S_0 (x_t)\\
&= \frac{G_F^2}{12 \pi^2} A^2 \lambda^4 M_W^2 f_{B_s}^2 m_{B_s} B_{B_s} S_0(x_t)~,}
containing two powers of $G_F$, where the loop function is
\begin{eqnarray}
S_0 (x_t) = \frac{4 x_t - 11 x_t^2 + x_t^3}{4(1-x_t)^2} - \frac{3 x_t^3 \log x_t}{2(1-x_t)^3} ,
\end{eqnarray}
for $x_t = m_t^2 / M_W^2$ and $\eta_{2B} = 0.551$. With these results we obtain \cite{Barger:2004qc}
\begin{eqnarray}
h_s = 3.858 \times 10^5 (\rho_{L}^{sb})^2 ,
\end{eqnarray}
where $\rho_L^{sb}$ is defined as
\begin{eqnarray}
\rho_L^{sb} \equiv \left|\frac{g''}{g_1} \frac{M_Z}{M_{Z'}} B_{23}^{d_L}\right|
= \left|\frac{g''}{g_1} \frac{M_Z}{M_{Z'}} (x-1) C_{q_L} V_{tb} V_{ts}^{\ast}\right|
\end{eqnarray}
which is constrained by the experimental results on $\Delta M_s$ of Eq. (\ref{eq:DMexp}),  $\Delta \Gamma_s$ of Eq. (\ref{eq:deltagam}), and $S_{\psi \phi}$ of Eq. (\ref{eq:spp}). The enhancement of the same-sign dimuon asymmetry by $\Gamma_{12}^s$ is also constrained by the other experimental measurement of $\Delta \Gamma_s$ and the indirect CP violation from $B_s \to J/\psi \phi$ decays such that \cite{Barberio:2008fa}
\begin{eqnarray}
\Delta \Gamma_s &=& \pm (0.154^{+0.054}_{-0.070})~ \text{ps}^{-1}
\label{eq:deltagam} \\
S_{\psi \phi} &=& -\sin\phi^s = - 0.77 ^{+0.29}_{-0.37}~~ \text{or}~ -2.36^{+0.37}_{-0.29}\,, \label{eq:spp}
\end{eqnarray}
where we assumed Arg.($-V_{ts}V_{tb}^{\ast}/V_{cs} V_{cb}^{\ast}$) is highly suppressed. The allowed parameter set $\sigma_s - h_s$ is shown in Fig. \ref{fig:hsigsbound}. As a result, for $h_s \lesssim 0.3$, \ie, $\rho_L^{sb} \lesssim 8 \times 10^{-4}$, there is no constraint on $\sigma_s$.
In this case, $|(x-1)C_{q_L}| \lesssim 2 \times 10^{-2}$. Even though we adopt a larger $h_s \lesssim 2.5$ depending on the value of $\sigma_s$, the value of $\rho_L^{sb}$ is at most $2.55 \times 10^{-3}$ so that $|(x-1)C_{q_L}| < 6.03 \times 10^{-2}$. Therefore, either $C_{q_L}$ is very small or $x$ is not much deviated from the value $x=1$. To obtain sizable $\Gamma_{12}^{s \rm NP}$ comparable to $\Gamma_{12}^{s \rm SM}$, the latter must be considered since $\Gamma_{12}^{s \rm NP}$ is proportional to $(B_{ij}^{\psi_{L,R}})^2 \propto \mathcal{O}(x^2 C_{q_L}^2)$. 

\begin{figure}
\begin{center}
\epsfig{figure=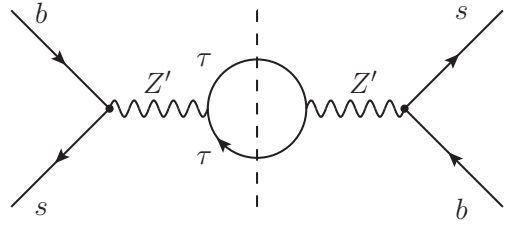,width=7cm}
\end{center}
\caption{The one-loop induced $B_s - \overline{B}_s$ mixing. The absorptive part indicates the $B_s (\overline{B}_s) \to \tau^- \tau^{+}$  decay, which provides a NP contribution to $\Gamma_{12}^s$.}
\label{fig:zprime2}
\end{figure}

\begin{figure}
\begin{center}
\epsfig{figure=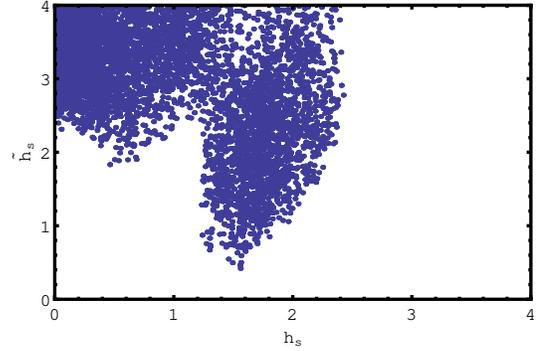,width=7cm}
\end{center}
\caption{The allowed parameter region in the $\tilde{h}_s - h_s$ plane from the experimental bound on $\Delta M_s$ (Eq. (\ref{eq:DMexp})), $\Delta \Gamma_s$ (Eq. (\ref{eq:deltagam})), and $S_{\psi \phi}$ (Eq. (\ref{eq:spp})). The parameter $\tilde{h}_s \gtrsim 2.5$ is favored for $h_s \lesssim 0.3$. $\tilde{h}_s$ could be lowered for $\sigma_s \sim \pi/2$ and $h_s \sim 1.5 - 2.0$.}
\label{fig:hsrbound}
\end{figure}
The width difference $\Gamma_{12}^s$ can be obtained from the absorptive part of the one-loop induced mixing of Fig. \ref{fig:zprime2}. From the calculations of $\Gamma_{12}^q$ of previous works \cite{Golowich:2006gq,Chen:2007dg,Alok:2010ij}, we obtain
\begin{eqnarray}
\tilde{h}_s \sim \left( \frac{\rho_L^{sb}}{|V_{cb}|} \right)^2 \left( B_{\tau \tau}^{\ell_L} + B_{\tau \tau}^{e_R} \right)^2~.
\end{eqnarray}
We assumed that the contribution to $\Gamma_{12}^{s \rm NP}$ is dominated by the $\tau$ loop insertion. This assumption is reasonable compared to the contribution by light quarks. According to Table. \ref{table:charge}, we obtain $\left( B_{\tau \tau}^{\ell_L} + B_{\tau \tau}^{e_R} \right)^2 = 9 (2x+1)^2 C_{q_L}^2$. For the loop contribution by light quarks ($u$, $d$, and $s$ quarks), $\Gamma_{12}^{s \rm NP}$ is proportional to $(B_{ij}^{\psi_{L,R}})^2$ which is at most $(5 x^2 + 8 x + 4) C_{q_L}^2$. If $x<-0.7$ or $x>-0.1$, the contribution to the $\tau^-\tau^+$ mode compared to the light quark modes is greater than two. This ratio is about five if $|x-1| \ll 1$. This situation is the same for $\Gamma$ of $B$ meson decays since it is also proportional to $(B_{i,j}^{\psi_{L,R}})^2$. Therefore, the contributions by the dangerous operators $(\bar{b}s)(\bar{\psi}\psi)$ for $\psi = u, d, s$ quarks to $\Gamma_{12}^{s \rm NP}$ and the decay rate ($\Gamma$) of $B$ mesons can be suppressed compared to the $\tau$ contribution. If this suppression is not enough, we may consider an additional flavor non-universal U(1)$'$ charge assignments to the right-handed quarks.

For the lepton contribution for $\psi = e, \mu$, however, the situation is different since U(1)$'$ charges are the same as that of $\tau$. The parameter $B_{\tau \tau}^{\ell_L}$ is constrained by the search of Br.$\,(B_s \to \mu^- \mu^+)$ at the Tevatron and the inclusive $b \to s \ell^{-} \ell^{+}$ decays \cite{Cheung:2006tm}. The upper bounds on the leptonic decay modes are Br.$\,(B_s\to\tau^-\tau^+) < 0.05$ and  Br.$\,(B_s\to\mu^-\mu^+)< 4.7\times 10^{-8}$ \cite{PData08}. Therefore, we have to consider the flavor non-universal U(1)$'$ charge assignment to the leptons also. It is not difficult to obtain this with an anomaly free U(1)$'$ by assuming exotic heavy leptons. For example, as done in Eq. (\ref{eq:qnonuniv}) for quarks, we can give flavor non-universal charges to leptons. If we assign the same $y$ for the left and right handed leptons as
\dis{ \epsilon_{\ell_L}^{(2)} =
  & C_{\ell}\left ( \begin{array}{ccc}
   y & 0 & 0  \\
   0 & y & 0 \\
  0 & 0 & 1
  \end{array} \right)~,\label{eq:lnonuniv}
}
we constrain $y\lesssim 4.34\times 10^{-3}$ to satisfy the above  Br.$\,(B_s\to\mu^-\mu^+)$ bound. For this parameter $y$, we can easily make a model to cancel all gauge anomalies, which is not exposed here.

Considering the constraints of Eq. (\ref{eq:DMexp}), (\ref{eq:deltagam}), and (\ref{eq:spp}), it is possible to obtain $\mathcal{O}(1)$ values of $\tilde{h}_s$ to explain the observed dimuon asymmetry.  Such parameter space is shown in Fig. \ref{fig:hsrbound} in the $\tilde{h}_s - h_s$ plane.

\begin{figure}
\subfigure[~$b \to s \gamma$ due to $Z'$ with $\tau$-loop exchange]{
\epsfig{figure=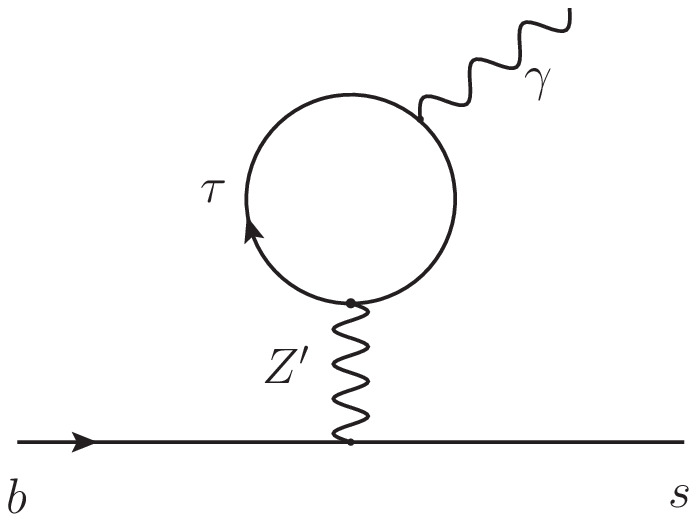,width=0.4\textwidth,angle=0}}
\quad
\subfigure[~$b \to s \gamma$ similar to the SM case]{
\epsfig{figure=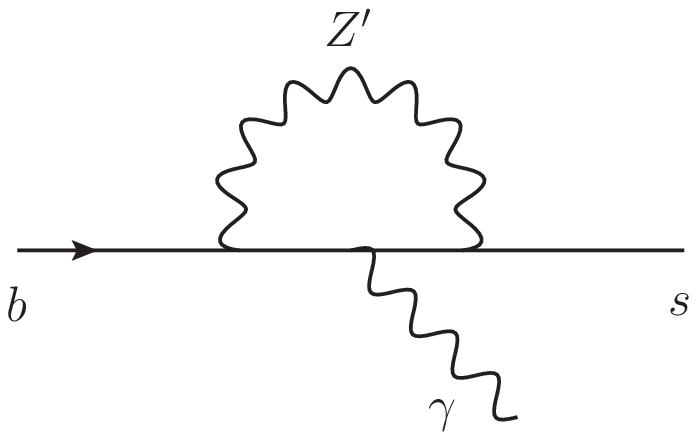,width=0.4\textwidth,angle=0}}
\caption{The leading order NP contributions by our $Z'$ model to $b \to s \gamma$ decay are depicted. The contribution due to the $(\bar{s}b)(\bar{\tau} \tau)_{V,A}$ operator through $Z'$ exchange is shown in (a), while another NP contribution in $Z'$ models is shown in (b).}
\label{fig:bsr}
\end{figure}

Now, we are required to consider the experimental constraint from the $b \to s \gamma$ decay which induces the $B \to X_s \gamma$ decays. The recent experimental measurements provide $Br.(B \to X_s \gamma) = (3.52 \pm 0.25) \times 10^{-4}$ \cite{Barberio:2008fa} which is consistent with the theoretical prediction $Br.(B \to X_s \gamma)^{\rm th} = (3.15 \pm 0.23) \times 10^{-4}$ \cite{Misiak:2006zs}. Therefore, the allowed contribution due to the NP effects is only within $\mathcal{O}(10 \%)$ of the SM contribution, which makes the NP model construction difficult. The NP operators of the form $\bar{s}b \bar{\psi} \psi$ where $\psi$ is a SM fermion contribute to $B \to X_s \gamma$ through the mixing with the operator of the form ($\bar{s} \sigma^{\mu \nu} b F_{\mu \nu}$). Focussing on the $(\bar{s} b)(\bar{\tau} \tau)_{V,A}$ operator, as analyzed in \cite{Bauer:2010dga}, our leading order NP contribution to $B \to X_s \gamma$ can be compared to the SM contribution such that
\begin{eqnarray}
\frac{{\rm Br.}(B \to X_s \gamma)^{\rm NP}}{
{\rm Br.}(B \to X_s \gamma)^{\rm SM}} \sim 2 \frac{\alpha}{\alpha_s} \sqrt{\tilde{h}_s} \sim 0.07~, \label{eq:bsgam}
\end{eqnarray}
for $\tilde{h}_s=1$ which is in the allowed region of Fig. \ref{fig:hsrbound}. Therefore, it is possible to obtain viable parameter space which does not suffer from the experimental constraints of $b \to s \gamma$ decay. (If the value of $h_s$ is chosen to be around 0.3 or smaller, the allowed value of $\tilde{h}_s$ is $\tilde{h}_s \gtrsim 2$. Then, the ratio (\ref{eq:bsgam}) is around $0.17$ which is somewhat big. However, it is marginally within $\mathcal{O}(10 \%)$ contribution.) The leading order NP contribution by our $Z'$ model to $b \to s \gamma$ decay is depicted in Fig. \ref{fig:bsr}. The contribution due to the $B_s \to \tau \tau$ through $Z'$ exchange is shown in Fig. \ref{fig:bsr} (a), while another NP contribution in $Z'$ models is shown in (b). One might also consider the contribution by $b \to s \gamma \nu \bar{\nu}$ through the tree level $Z'$ exchange, however, it is a four-body decay process which is suppressed by phase factor $(1/4\pi)^4$ so that a subdominant contribution. Consequently, our $Z'$ model considering the enhancement of $\Gamma_{12}^{s\, \rm NP}$ due to the $B_s \to \tau^- \tau^+$ decay is safe from $b \to s \gamma$.


\section{Conclusions}
\label{sec:conclusions}

We have analyzed various NP models to explain the large same-sign dimuon asymmetry observed by the D0 experiment. The observed central value has about $3.2 \sigma$ discrepancy from the SM prediction. Combining it with the other experimental results, the total average value is shifted to about $2.5 \sigma$ discrepancy. The MSSM with minimal flavor violation and multi Higgs models cannot provide such a large asymmetry due to their loop effects or Yukawa suppressions. On the other hand, a $Z'$ model with extra U(1)$'$ gauge symmetry can accommodate the observed result.   For a theoretically viable $Z'$ model, we considered the case that the U(1)$'$ quark charges are assigned to be flavor non-universal, which is in fact a general phenomenon with an extra U(1)$'$. To preserve the extra non-anomalous U(1)$'$  symmetry at high energy, SU(2)$_L$ singlet exotic heavy quarks of mass above 1 TeV and the SM gauge singlet scalars are introduced. The other experimental results such as $B \to X_s \gamma$, $\Delta M_s, \Delta \Gamma_s$, and $S_{\psi \phi}$ are also considered in our $Z'$ model. To satisfy the additional constraints such as $B_s \to \mu^- \mu^+$, a flavor non-universal U(1)$'$ charge assignment to leptons or right-handed quarks.
is needed, which is not difficult to obtain with anomaly free U(1)$'$. In conclusion, we presented the case of allowing a large NP CP violation with an extra $Z'$, with the parameter region consistent with the various present experimental bounds.

\acknowledgments{J.E.K. thanks W. Grimus, G. Isidori, and H. Neufeld for useful discussions. This work is supported in part by the Korea Research Foundation, Grant No. KRF-2005-084-C00001, and  MSS and SS are supported in addition by Grant No. KRF-2008-313-C00162 and the FPRD of the BK21 program.}

\vskip 0.5cm

\appendix

\section{The down type quark sector in the dodeca model}
\label{appendix:dodeca}

To make ($bs$) FCNC, we consider different matrix from original dodeca model  \cite{KimSeo1}.
The unitary matrices diagonalizing the quark mass matrices are given by
\dis{
&U_u = \left ( \begin{array}{ccc}
   \frac{1}{\sqrt{2}} & -\frac{1}{\sqrt{2}}\mathrm{e}^{i \phi} & 0  \\
   \frac{1}{\sqrt{2}} \mathrm{e}^{-i \phi} & \frac{1}{\sqrt{2}} & 0 \\
   0 & 0 & 1
  \end{array} \right)\\
 & U_d =  \left ( \begin{array}{ccc}
   \frac{1}{\sqrt{2}} & -\frac{1}{\sqrt{2}} & 0  \\
   \frac{1}{\sqrt{2}} & \frac{1}{\sqrt{2}} & 0 \\
   0 & 0 & 1
  \end{array} \right)
  }

To obtain a CP violation effect in the $B_s$ system in the dodeca model, the loop diagrams similar to Fig. \ref{fig2:TwoHiggsCP} must be considered. Namely, the entries of (23) element and/or (32) element of $V_{\rm CKM}$ are needed. As pointed out in \cite{KimSeo1}, this appears as a small correction in $U_d$,
\dis{
 U_d^\prime &= \left ( \begin{array}{ccc}
   1 & 0 & \epsilon \mathrm{sin}\frac{\phi}{2}  \\
  0  & 1 & - i\epsilon \mathrm{cos}\frac{\phi}{2} \\
   -\epsilon \mathrm{sin}\frac{\phi}{2} & -i\epsilon \mathrm{cos}\frac{\phi}{2} & 1
  \end{array} \right) \left ( \begin{array}{ccc}
   \frac{1}{\sqrt{2}} & -\frac{1}{\sqrt{2}} & 0  \\
   \frac{1}{\sqrt{2}} & \frac{1}{\sqrt{2}} & 0 \\
   0 & 0 & 1
  \end{array} \right)
  \\
  & =   \left ( \begin{array}{ccc}
   \frac{1}{\sqrt2} & -\frac{1}{\sqrt2} &  \epsilon \sin \frac{\phi}{2}  \\
   \frac{1}{\sqrt2} & \frac{1}{\sqrt2} &  -i \epsilon \cos \frac{\phi}{2} \\
  -\epsilon \frac{i}{\sqrt2} \mathrm{e}^{-i \frac{\phi}{2}} & -\epsilon \frac{i}{\sqrt2} \mathrm{e}^{i \frac{\phi}{2}}& 1
  \end{array} \right)
 }
The phase $\phi$ in the up type quark sector appears here in the down type quark sector. Basically, generation of O($\epsilon$) term is a loop effect as pointed out in Ref. \cite{KimSeo1}.  With this corrected matrix, the CKM matrix is given by
\dis{
V_{\rm CKM} = \left ( \begin{array}{ccc}
   \mathrm{e}^{i \frac{\phi}{2}} \cos \frac{\phi}{2} & -i \mathrm{e}^{i \frac{\phi}{2}} \sin \frac{\phi}{2} & 0 \\
  -i \mathrm{e}^{-i \frac{\phi}{2}} \sin \frac{\phi}{2} & \mathrm{e}^{-i \frac{\phi}{2}} \cos \frac{\phi}{2} & i \epsilon  \mathrm{e}^{-i \frac{\phi}{2}} \\
  \epsilon \sin \frac{\phi}{2}  & i \epsilon \cos \frac{\phi}{2} & 1
  \end{array} \right)\label{eq:VCKMdodeca}
}
For $ \phi = \pi/6 $, one should consider more corrections toward a more realistic $V_{\rm CKM}$, but here we omit this elaboration since it is not essential for our $\bar{s}b$ coupling. $\epsilon$ is of order 0.04 to make an agreement with the measured CKM matrix element.

The $U_d^\prime$ diagonalizes the mass matrix of the form
\begin{widetext}
\dis{
\left ( \begin{array}{ccc}
  \frac{1}{2} \Sigma_{12},  & \frac{1}{2} \Delta_{21}, & i\frac{\epsilon}{\sqrt2} (\Delta_{32} \cos \frac{\phi}{2} + i\Delta_{31} \sin \frac{\phi}{2}) \\
 \frac{1}{2} \Delta_{21},  & \frac{1}{2} \Sigma_{12}, & i\frac{\epsilon}{\sqrt2} (\Delta_{32} \cos \frac{\phi}{2} - i\Delta_{31} \sin \frac{\phi}{2})  \\
  -i\frac{\epsilon}{\sqrt2} (\Delta_{32} \cos \frac{\phi}{2} - i\Delta_{31} \sin \frac{\phi}{2}) ,  & -i\frac{\epsilon}{\sqrt2} (\Delta_{32} \cos \frac{\phi}{2} + i\Delta_{31} \sin \frac{\phi}{2}),  & m_3
  \end{array} \right)\label{eq:VRHd}
  }
where $\Delta_{21}=m_2-m_1,\Delta_{31}= m_3-m_1, \Delta_{32}=m_3-m_2, \Sigma_{12}=m_1+m_2$ and $\Sigma_{12}$, and
$m_1, m_2, \mathrm{and}~m_3$ are the eigenvalues after diagonalization. Note that $w_d=\frac{1}{2}(m_1+m_2) =\frac12\Sigma_{12},z_d=\frac{1}{2} (-m_1+m_2)=\frac12 \Delta_{21}$ and $x_d=m_3$. Using Eq. (\ref{eq:VRHd}), one can obtain the O($\epsilon$) Higgs couplings to the mass eigenstate $d$-type quarks.

In the mass eigenstate basis, this indicates that FCNC in down type quark sector is read as

  \dis{& U_d^{\dagger} H^{(d)} V_d
  \\
  & = \left ( \begin{array}{ccc}
  y_5^d H^{\prime}-y_4^d H_0^{\prime} &  \frac{1}{2}y_5^d (H_2^{\prime} - H_1^{\prime}) & -i\frac{\epsilon}{2} y_5^d (H_1^{\prime} - H_2^{\prime})\cos \frac{\phi}{2}\\
\frac{1}{2}y_5^d (H_2^{\prime} - H_1^{\prime}) &   y_5^d H^{\prime}+y_4^d H_0^{\prime} & \frac{\epsilon}{2} y_5^d (H_1^{\prime} - H_2^{\prime})\sin \frac{\phi}{2}\\
   i\frac{\epsilon}{2} y_5^d (H_1^{\prime} - H_2^{\prime})\cos \frac{\phi}{2} & \frac{\epsilon}{2} y_5^d (H_1^{\prime} - H_2^{\prime})\sin \frac{\phi}{2}  & y_1^d H_0
  \end{array} \right) }
\end{widetext}

The contribution to $\Gamma_{12}^s$ is given by
  \dis{\Gamma_{12}^{s \rm NP} \sim \vert f^{(d)} \vert ^4 \frac{\vert \mathrm{cos}\alpha^I_{22} \mathrm{cos}\alpha^I_{23} \vert ^2}{M^4_{\rho_I^{(d)+}}} M_{B^0_s}^5
  =
  \frac{\vert f^{(d)I}_{22} f^{(d)I}_{23} \vert ^2}{M^4_{\rho_I^{(d)+}}}
 M_{B^0_s}^5~,
   }
which involves the Yukawa couplings. Assuming that VEVs take the same value around $v= 250$GeV, we obtain

 \dis{&f^{(d)I}_{23} \simeq \frac{\epsilon}{\sqrt2 v}m_3 \mathrm{sin}\frac{\phi}{2}
 ,}
 and
  \dis{f^{(d)I}_{22} =  \simeq \frac{1}{2 v}m_2~. }


\vskip 0.5cm

\end{document}